\documentclass[reprint, superscriptaddress, amsmath, amssymb, aps]{revtex4-2}

\usepackage{graphicx}
\usepackage{dcolumn}
\usepackage{bm}
\usepackage{xcolor}
\usepackage{mathtools}
\usepackage{physics}
\usepackage[normalem]{ulem}
\usepackage{lineno}

\usepackage{hyperref} 
\hypersetup{colorlinks=true, citecolor=blue}

\begin{document}

\title{Statistics of Complex Wigner Time Delays as a counter of S-matrix poles: Theory and Experiment}

\author{Lei Chen}
 \email{LChen95@umd.edu}
 \affiliation{Quantum Materials Center, Department of Physics, University of Maryland, College Park, MD 20742, USA}
 \affiliation{Department of Electrical and Computer Engineering, University of Maryland, College Park, MD 20742, USA}

\author{Steven M. Anlage}
 \email{anlage@umd.edu}
 \affiliation{Quantum Materials Center, Department of Physics, University of Maryland, College Park, MD 20742, USA}
 \affiliation{Department of Electrical and Computer Engineering, University of Maryland, College Park, MD 20742, USA}

\author{Yan V. Fyodorov}
 \affiliation{Department of Mathematics, King’s College London, London WC26 2LS, United Kingdom}
 \affiliation{L. D. Landau Institute for Theoretical Physics, Semenova 1a, 142432 Chernogolovka, Russia}
 
\date{\today}
 
\begin{abstract}
We study the statistical properties of the complex generalization of Wigner time delay $\tau_\text{W}$ for sub-unitary wave chaotic scattering systems.  We first demonstrate theoretically that the mean value of the $\text{Re}[\tau_\text{W}]$ distribution function for a system with uniform absorption strength $\eta$ is equal to the fraction of scattering matrix poles with imaginary parts exceeding $\eta$.  The theory is tested experimentally with an ensemble of microwave graphs with either one or two scattering channels, and showing broken time-reversal invariance and variable uniform attenuation.  The experimental results are in excellent agreement with the developed theory.  The tails of the distributions of both real and imaginary time delay are measured and are also found to agree with theory.  The results are applicable to any practical realization of a wave chaotic scattering system in the short-wavelength limit.
\end{abstract}

\maketitle

\emph{Introduction.}  
In this paper we are concerned with the general scattering properties of complex systems, namely finite-size wave systems with one or more channels connected to asymptotic states outside of the scattering domain.  The scattering system is complex in the sense that classical ray trajectories will undergo chaotic scattering when propagating inside the closed system.  We focus on the properties of the energy-dependent scattering matrix of the system, defined via the linear relationship between the outgoing $\ket{\psi_\text{out}}$ and incoming wave amplitudes $\ket{\psi_\text{in}}$ on the $M$ coupled channels  as $\ket{\psi_\text{out}}=S\ket{\psi_\text{in}}$.  In the short wavelength limit the complex $M\times M$ scattering matrix $S(E)$ is a strongly fluctuating function of energy $E$ (or, equivalently, the frequency $\omega$) of the incoming waves, as well as specific system details. Those parts of the fluctuations which reflect long-time behavior are controlled by the high density of $S$-matrix poles, or resonances, having their origin at eigenfrequencies (modes) of closed counterparts of the scattering systems. At energy scales comparable to the mean separation $\Delta$ between the neighboring eigenfrequencies, the properties of the scattering matrix are largely universal, and depend on very few system-specific parameters. The ensuing statistical characteristics of the $S$-matrix have been very successfully studied theoretically over the last 3 decades using methods of Random Matrix Theory (RMT) \cite{VWZ85,MPS85,SokZel89,Fyodorov97,FyoSavSomRev05,MRW2010,FSav11,Nock14,Schomerus2015}.  
 
The scattering matrix can be characterized by the distribution of  poles and associated zeros in the complex energy plane, which are most clearly seen when one addresses its determinant.  In the unitary (zero loss) limit, the poles and zeros of the determinant form complex conjugate pairs across the real axis in the energy plane.  In the presence of any loss, the poles and zeros are no longer complex conjugates, but if the loss is spatially-uniform their positions are still simply related by a uniform shift. This is no longer the case for spatially-localized losses, with poles and zeros migrating in a complicated way to new locations, subject to certain constraints. For a passive lossy system the poles always remain in the lower half of the complex energy plane, while the zeros can freely move between the two sides of the real axis. Among other things, rising recent interest in characterizing $S$-matrix complex zeros, as well as their manifestation in physical observables, is strongly motivated by the phenomenon of coherent perfect absorption \cite{Baran17}, see \cite{Li2017,Fyodorov2017,Fyo2019,OsmanFyo20,Chen2021gen} and references therein.
 
One quantity which is closely related to resonances is known to be the Wigner time delay $\tau_\text{W}$.
In its traditional definition \cite{Wigner55,Smith60}  for unitary, flux conserving scattering systems the Wigner time delay is a real positive quantity measuring how long an excitation lingers in the scattering region before leaving through one of the $M$ channels.   Statistical fluctuations of  $\tau_\text{W}$ in  flux-conserving systems with no internal losses was the subject of a large number of theoretical works in the RMT context \cite{Lehmann95,FyodSomm96,Gopar96,Fyodorov97a,Brouwer97,Brouwer99,SFS01,OssFyo05,Kott05}, and more recently \cite{MezSimm13,TexMaj013,Cunden15,Texier16,Grabsch20}, as well in a semiclassical context in \cite{Kuipers14,Novaes15,Uzy17,Uzy18} and references therein.  In particular, for the one and two channel cases most relevant to this paper the distribution of $\tau_\text{W}$ is known explicitly for all symmetry classes, $\beta=1$, 2 and 4 \cite{SFS01}.
 
Experimental work on time delays in wave chaotic billiard systems was pioneered by Doron, Smilansky and Frenkel in microwave billiards with uniform absorption \cite{Doron90}, where the relation between the Wigner time delays and the unitary deficit of the $S$-matrix has been explored.  Later experiments on time delay statistics were made by Genack and co-workers, who studied microwave pulse delay times through randomized dielectric scatterers \cite{Genack99,Genack03}.  The quantity studied in that case is a type of partial time delay associated with the complex transmission amplitude between channels \cite{Tigg99}, somewhat different from the Wigner time delay. In particular, contributions to the transmission time delay due to poles and zeros of the off-diagonal $S$-matrix entries have been identified \cite{Genack21}.
 
Despite strong interest in the standard Wigner time delay over  the years, its use for characterising statistics of $S$-matrix poles and zeros beyond the regime of well-resolved (isolated) resonances have been always problematic. In our recent paper \cite{Chen2021gen} we proposed a  complex generalization of the Wigner time delay $\tau_\text{W}$ as a tool for identifying the locations of the poles and zeros of the $S$-matrix. In its generalized version, the Wigner time delay in the presence of losses becomes a complex function of energy \cite{Chen2021gen} and reflects the phase and amplitude variation of the scattering matrix with energy.  Subsequently, we developed a method, both experimentally and theoretically, for exploiting the generalized complex Wigner time delays (CWTD) for identifying the locations of individual $S$-matrix poles $\mathcal{E}_n$ and zeros $z_n$ in the complex energy plane. The method has been implemented in the regime of well-resolved, isolated resonances, for systems with both localized and uniform sources of absorption. However, no statistical characterization of CWTD for large numbers of modes has been attempted.

In the present work we go at the next level in exploiting these ideas by revealing and analysing the statistical properties of generalized complex Wigner time delay in wave-chaotic scattering systems, including the most challenging regime of partly overlapping resonances.  The purpose of this paper is both to present new experimental results on statistics of complex Wigner time delays in wave chaotic systems and to provide a theoretical understanding of some features observed experimentally.  The results are applicable to any practical realization of a wave chaotic scattering system in the short-wavelength limit, including quantum wires and dots, acoustic and electromagnetic resonators, and quantum graphs.


To this end it is worth mentioning that one of the oldest yet useful facts about the standard Wigner time delay is that the mean of the $\tau_\text{W}$ distribution is simply related to the Heisenberg time $\tau_\text{H}$ of the system, $\langle \tau_\text{W}\rangle = 2\pi \hbar/M\Delta:=\tau_\text{H} /M$ \cite{Lyub77}. As such it is absolutely insensitive to the type of dynamics, chaotic versus integrable. More recently this property was put in a much wider context and tested experimentally \cite{Pierrat2014}.

In this paper we reveal that the mean value of $\text{Re}[\tau_\text{W}]$ of CWTD is, in striking contrast to the flux-conserving case, a much richer object and can be used to obtain nontrivial information about the distribution of the imaginary part of the poles of the $S$-matrix. For this we develop the corresponding theory for the mean values and compare to the experimentally observed evolution of distributions of real and imaginary parts of CWTD with uniform loss variation. 

\begin{figure*}[ht]
\includegraphics[width=\textwidth]{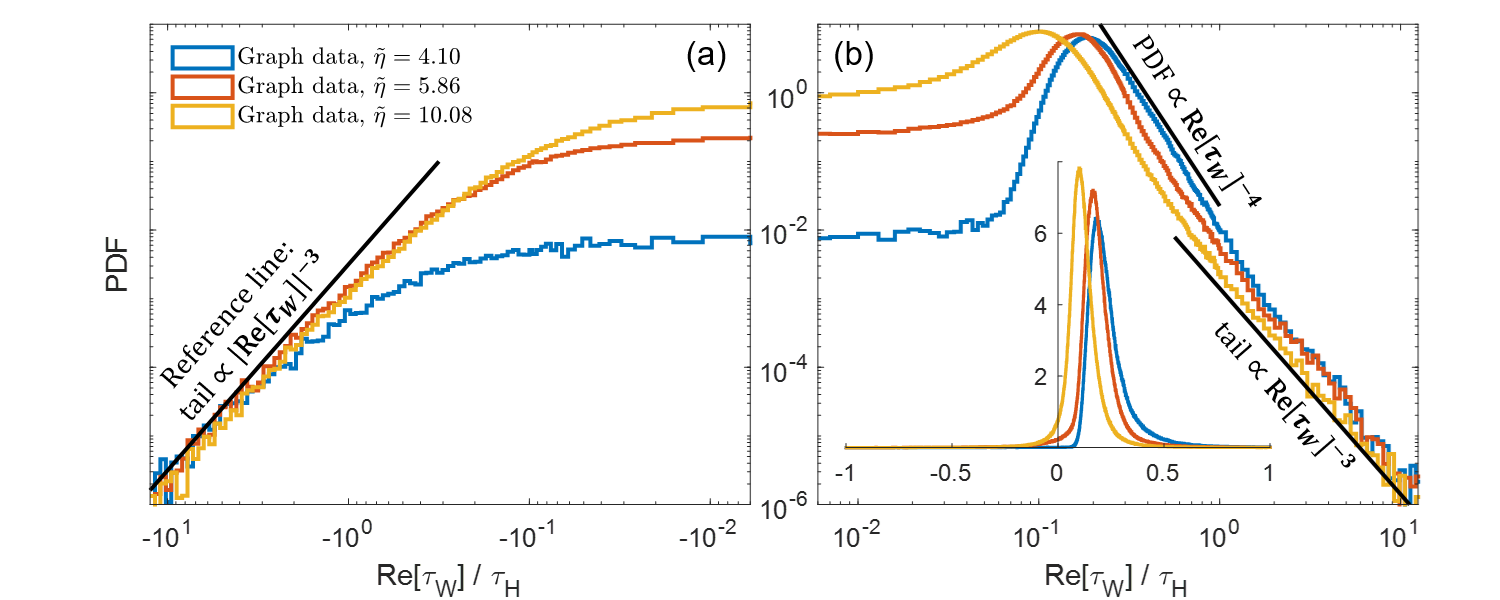}
\caption{Evolution of the PDF of measured $\text{Re}[\tau_\text{W}]$ with increasing uniform attenuation ($\tilde{\eta}$) from an ensemble of two-port ($M=2$) tetrahedral graph data with broken-TRI. (a) and (b) show the distributions of the negative and positive $\text{Re}[\tau_\text{W}]$ on a log-log scale for three values of uniform attenuation, respectively. Reference lines characterizing power-law behavior are added to the tails. Inset in (b) shows the distributions of $\text{Re}[\tau_\text{W}]$ on a linear scale for the same measured data.}
\label{fig1}
\end{figure*}

\emph{Theory.}
The appropriate theoretical framework for our  analysis is the so called effective Hamiltonian formalism for wave-chaotic scattering \cite{SokZel89,Fyodorov97,FSav11,kuhl13,Schomerus2015}. It starts with defining an $N \times N$ self-adjoint matrix Hamiltonian $H$ whose real eigenvalues are associated with eigenfrequencies of the closed system. Further defining $W$ to be an $N \times M$
matrix of coupling elements between the $N$ modes of $H$ and the $M$ scattering
channels, one builds the unitary $M\times M$ scattering matrix $S$ in the form:
\begin{align}
    \label{eq1}
    \mathcal{S}(E)=1_M - 2\pi i W^{\dag} \frac{1}{E-H+i\Gamma_W} W
\end{align}
where we defined $\Gamma_W = \pi WW^{\dag}$.
Note that in this approach the $S$-matrix poles $\mathcal{E}_n = E_n -i\Gamma_n$ (with $\Gamma_n>0$) are complex eigenvalues of the non-Hermitian effective Hamiltonian matrix ${\cal H}_\text{eff}=H-i\Gamma_W\ne {\cal H}_\text{eff}^{\dagger}$.

A standard way of incorporating the uniform absorption with strength $\eta$ is to replace $E\to E+i \eta$ in the $S$ matrix definition. Such an $S$-matrix becomes subunitary and we denote  $S(E + i\eta) := S_{\eta}(E)$. The determinant of $S_{\eta}(E)$ is
then given by
\begin{align}
    \label{eq2}
    \det S_{\eta}(E) &\coloneqq \det S(E+i\eta) \\
    \label{eq3}
    &= \frac{\det [E-H+i(\eta  - \Gamma_{W})]}
    {\det [E-H+i(\eta  + \Gamma_{W})]} \\
    \label{eq4}
    &= \prod_{n=1}^{N} \frac{E+i\eta - \mathcal{E}_n^*}{E+i\eta - \mathcal{E}_n},
\end{align}

Using the above the expression, the Wigner time delay can be very naturally extended to scattering systems with uniform absorption as suggested in \cite{Chen2021gen} by defining:

\begin{widetext}
\begin{align}
    \label{eq5}
    \tau_\text{W}(E;\eta) &\coloneqq \frac{-i}{M} \frac{\partial}{\partial E} \log \det S_{\eta}(E) \\
    \label{eq6}
    &= \text{Re}\ \tau_\text{W}(E;\eta) + i\text{Im}\ \tau_\text{W}(E;\eta), \\
    \label{eq7}
    \text{Re}\ \tau_\text{W}(E;\eta) &= \frac{1}{M} \sum_{n=1}^{N} \left[ \frac{\Gamma_n + \eta}{(E-E_n)^2 + (\Gamma_n + \eta)^2} - \frac{\eta -\Gamma_n }{(E-E_n)^2 + (\Gamma_n - \eta)^2} \right], \\
    \label{eq8}
    \text{Im}\ \tau_\text{W}(E;\eta) &= -\frac{1}{M} \sum_{n=1}^{N} \left[ \frac{4\eta \Gamma_n(E - E_n)}{[(E-E_n)^2 + (\Gamma_n - \eta)^2] [(E-E_n)^2 + (\Gamma_n + \eta)^2]}  \right]
\end{align}
\end{widetext}

For a wave-chaotic system the set of parameters $\Gamma_n,E_n$ (known as the {\it resonance widths} and {\it positions}, respectively) is generically random. Namely, even minute changes in microscopic shape characteristics of the system will drastically change the particular arrangement of $S$-matrix poles in the complex plane in systems which are otherwise macroscopically indistinguishable.  To study the associated statistics of CWTD most efficiently one may invoke the notion of an {\it ensemble} of such systems. As a result, both $\text{Re}[\tau_\text{W}]$ and $\text{Im}[\tau_\text{W}]$ at a given energy  will be distributed over a wide range of values. Alternatively, even in a single wave-chaotic system the CWTD will display considerable statistical fluctuations when sampled over an ensemble of different  {\it mesoscopic} energy intervals, see below and \cite{SuppMat} for more detailed discussion. Invoking the notion of spectral ergodicity one expects that in wave-chaotic systems the two types of ensembles should be equivalent. 

In the Supp. Mat. section I \cite{SuppMat} we analyse in much detail the mean value of the CWTD in systems with uniform absorption $\eta>0$. In contrast to the case of flux-conserving systems the mean of $\text{Re}[\tau_\text{W}]$ becomes highly nontrivial as it counts the number of $S$-matrix poles whose widths exceed the uniform absorption strength value.  In other words,

\begin{widetext}
\begin{align}
    \label{eq9}
    \langle \text{Re}[\tau_\text{W}(E;\eta)]\rangle_E = \frac{\tau_\text{H}}{M} \times \text{Prob} (\text{resonance\ widths} > \eta)
\end{align}
where we defined
\begin{align*}
    \text{Prob} (\text{resonance\ widths} > \eta) \coloneqq \frac{\#[\Gamma_n>\eta\ \text{such\ that}\ E_n\ \text{is\ inside}\ I_E]}{\text{total\ \#\ resonances\ inside}\ I_E}
\end{align*}
\end{widetext}
where $I_E$ is a mesoscopic energy scale that is defined to be much larger than the mean mode spacing $\Delta$ but small enough so that the interval has a roughly constant mode density.
Alternatively, invoking ergodicity, one may use the RMT for analysing the mean CWTD, which independently confirms Eq. (\ref{eq9}). Such analysis also predicts that $\langle \text{Im}[\tau_\text{W}(E,\eta)\rangle_E =0$, independent of $\eta$.  The distribution of imaginary parts $\Gamma_n$ of the $S$-matrix poles relevant for Eq. (\ref{eq9}) have been examined theoretically in the RMT framework \cite{Haake92,Fyod96,Somm99,FyodKhor99} and experimentally \cite{Kuhl08,Difa12,Bark13,Liu14,Gros14,DavyGenack18} by a number of groups. 
 
 \begin{figure}[ht]
\includegraphics[width=86mm]{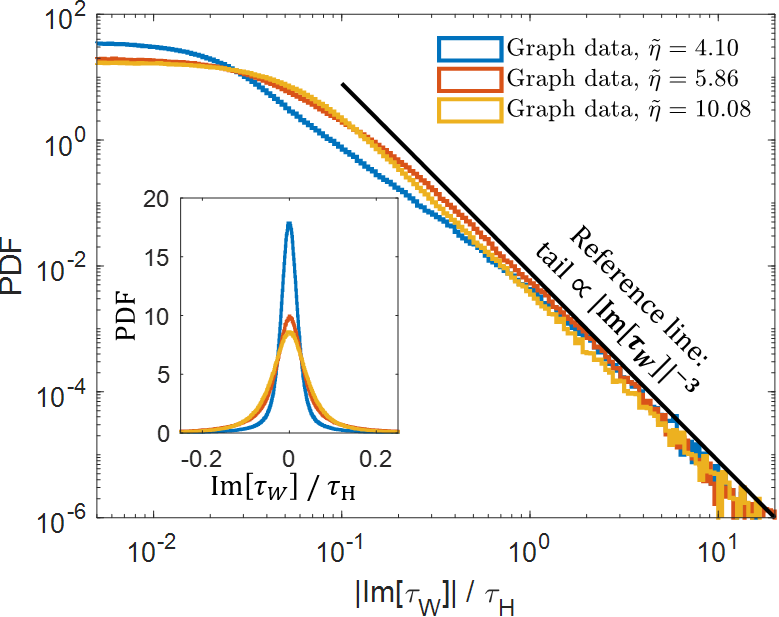}
\caption{Evolution of the PDF of measured $\text{Im}[\tau_\text{W}]$ with increasing uniform attenuation ($\tilde{\eta}$) from an ensemble of two-port ($M=2$) tetrahedral graph data with broken-TRI. The main figure shows a log-log plot of the PDF versus $|\text{Im}[\tau_\text{W}]|$ for three values of uniform attenuation in which the data for negative values has been folded over to the positive side.  A reference line is added to characterize the power-law tail.  Inset shows the distributions of $\text{Im}[\tau_\text{W}]$ on a linear scale for the same measured data.}
\label{fig2}
\end{figure}

\emph{Experiment.}  We test our theory by using an ensemble of tetrahedral microwave graphs with either $M = 1$ or $M = 2$ channels coupled to the outside world.  We focus on experiments involving microwave graphs \cite{Hul04,Lawn08,Hul12,Chen2020} for a number of reasons.  Microwave graphs have a number of advantages for wave chaotic statistical studies: one can precisely vary the uniform loss and the lumped loss over a wide range; one can work in either the time-reversal invariant (TRI) or broken-TRI regimes; one can gather very good statistics with a large ensemble of graphs; one can change $M$ from 1 to 2 in a convenient manner; one can vary both the (energy-independent) mode density and loss to go from the limit of isolated modes to strongly overlapping modes.  The disadvantages of graphs for statistical studies include significant reflections at nodes, which can create trapped modes on the bonds \cite{Fu17}, and the appearance of short periodic orbits in cyclic graphs \cite{Dietz17}.  As a result of these limitations some statistical properties show deviations from RMT predictions. 
	
\begin{figure}[ht]
\includegraphics[width=86mm]{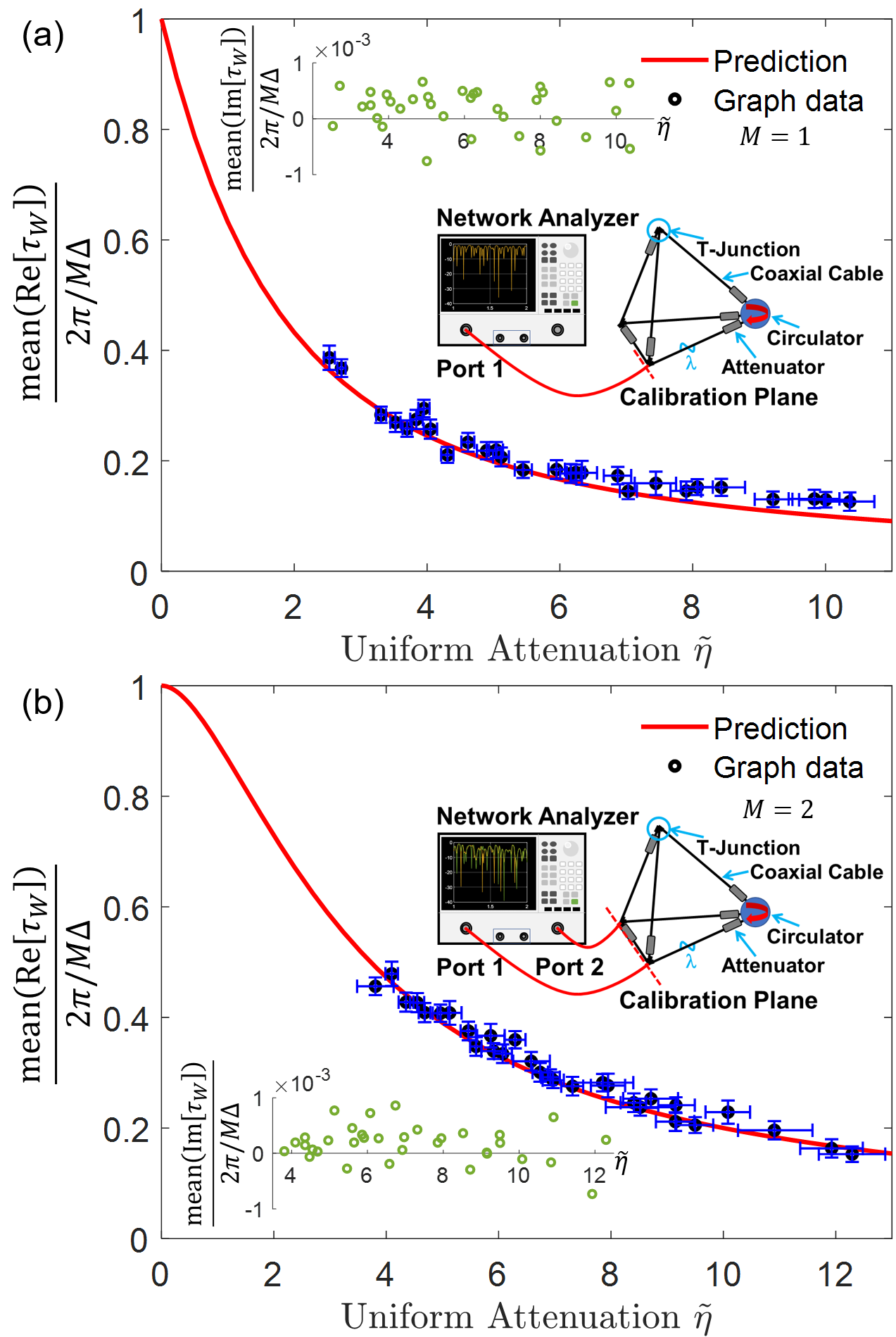}
\caption{Mean of the $\text{Re}[\tau_\text{W}]$ as a function of uniform attenuation $\tilde{\eta}$ evaluated using tetrahedral graph data with broken-TRI for both one- and two-port configurations. (a) shows the one-port experimental data (black circles) compared with theory (red line). (b) shows the two-port experimental data (black circles) compared with theory (red line). A detailed discussion about the estimated error bars (blue) can be found in Supp. Mat. section V. \cite{SuppMat} Insets show the mean of the $\text{Im}[\tau_\text{W}]$ (green circles) as a function of uniform attenuation $\tilde{\eta}$ evaluated using the same datasets for the one- and two-port configurations, respectively.}
\label{fig3}
\end{figure}

The microwave graphs are constructed with coaxial cables with center conductors of diameter 0.036 in (0.92 mm) made with silver plated copper clad steel, and outer shield of diameter 0.117 in (2.98 mm) made with a copper-tin composite.  An ensemble of microwave graphs is created by choosing 6 out of 9 cables with different incommensurate lengths (for a total of $\binom{9}{6}=84$ realizations) and creating uniquely different tetrahedral graphs. The scattering matrix of the 1 and 2-port graphs are measured with a calibrated Agilent PNA-X N5242A Network Analyzer (see insets of Fig. \ref{fig3}) over the frequency range from 1 to 12.4 GHz, which includes about 250 modes in a typical realization of the ensemble. The graphs are measured with a finite coupling of $g_a$ which varies from 1.06 to 1.80 as a function of frequency, where $g_a=\frac{2}{T_a}-1$ and $T_a=1-|S_{\text{rad}}|^2$ is the transparency of the graph to the scattering channel $a$ determined by the value of the radiation $S$-matrix. \footnote{The radiation $S_{\text{rad}}$ is measured when the graph is replaced by $50\ \Omega$ loads connected to the three output connectors of each node attached to the network analyzer test cables.}  The effects of the coupling are then removed through application of the Random Coupling Model (RCM) normalization process \cite{Hemmady05,Henry1,Henry2,Grad14}. This is equivalent to creating an ensemble of data with perfect coupling, $g_a=1$ and $T_a=1$ for all frequencies, ports, and realizations. 

Time-reversal invariance was broken in the graph by means of one of 4 different microwave circulators \cite{awniczak2010} operating in partially overlapping frequency ranges going from 1 to 12.4 GHz (see Supp. Mat. section VI \cite{SuppMat}).  The complex Wigner time delay $\tau_\text{W}$ is calculated using the RCM-normalized scattering matrix $S$ as in Eq. (\ref{eq5}), and the statistics of the real and imaginary parts are compiled based on realization averaging and frequency averaging in a given frequency band.  Each frequency band is chosen to have a large number of modes (approximately 40) but small enough so that the uniform attenuation value is approximately constant. A total of 84 realizations of the graphs were created, and the data was broken into 7 frequency bands of approximately equal attenuation.  The overall level of attenuation was varied by adding identical fixed microwave attenuators to each of the 6 bonds of the tetrahedral graphs \cite{Lawn19}.  The attenuator values chosen were 0.5, 1 and 2 dB.  

\emph{Comparison of Theory and Experiments.}  Our prior work showed that CWTD varied systematically as a function of energy/frequency for an isolated mode of a microwave graph \cite{Chen2021gen}.  The real and imaginary parts of $\tau_\text{W}$ take on both positive and negative values.  We now consider an ensemble of graphs and examine the distribution of these values taken over many realizations and modes.  We first examine the evolution of the PDF of $\text{Re}[\tau_\text{W}]$ (inset of Fig. \ref{fig1}(b)) and $\text{Im}[\tau_\text{W}]$ (inset of Fig. \ref{fig2}) with increasing uniform (normalized) attenuation $\tilde{\eta}$.  The uniform attenuation is quantified from the experiment as $\tilde{\eta} = \frac{2\pi}{\Delta} \eta = 4\pi\alpha$, where $\alpha=\delta f_\text{3dB}/\Delta_f$, $\delta f_\text{3dB}$ is the typical 3-dB bandwidth of the modes and $\Delta_f$ is the mean frequency spacing of the modes \cite{Hemmady06}.

Fig. \ref{fig1} shows that as the uniform attenuation ($\tilde{\eta}$) of the graphs increases, the peak of the $\text{Re}[\tau_\text{W}]$ distribution shifts to lower values. Furthermore, Fig. \ref{fig1}(a) shows that $\text{Re}[\tau_\text{W}]$ acquires more negative values as the attenuation increases. Both Fig. \ref{fig1}(a) and (b) demonstrate that the PDF of $\text{Re}[\tau_\text{W}]$ exhibits power-law tails on both the negative and positive sides, respectively. The positive-side PDFs shown in Fig. \ref{fig1}(b) have different power-law behaviors for different ranges of $\text{Re}[\tau_\text{W}]$, which is further explained theoretically in the Supp. Mat. section II \cite{SuppMat}. Fig. \ref{fig2} shows the PDF of $|\text{Im}[\tau_\text{W}]|$ on both linear and log-log scales for the same values of uniform attenuation.  We find that the $\text{Im}[\tau_\text{W}]$ distribution is symmetric about zero to very good approximation.  Once again a power-law behavior of the tails of the distribution is evident.  The inset in Fig. \ref{fig2} demonstrates that these distributions become lower and broader as the uniform attenuation ($\tilde{\eta}$) of the graph system increases.
  
Figure \ref{fig3} shows a plot of the Mean$(\text{Re}[\tau_\text{W}])$ vs. uniform attenuation ($\tilde{\eta}$) in ensembles of microwave graphs for both (a)  $M = 1$ and (b) $M = 2$ ports.  The black circles represent the data taken on an ensemble of microwave graphs with constant $\tilde{\eta}$.  The red line is an evaluation of the relation Eq. (\ref{eq9}) above, based on the analytical prediction for the $P(\Gamma_n)$ distribution for the a) $M=1$ and b) $M=2$ cases, both with perfect coupling ($g=1$) \cite{Fyod96,Fyodorov97}.  Note that the distribution of $\Gamma_n$ for $M=1$ is very different from the multi-ports cases (see Fig. S3 in the Supp. Mat. \cite{SuppMat}). Nevertheless there is excellent agreement between data and theory over the entire experimentally accessible range of uniform attenuation values for both 1-port and 2-port graphs.  We can conclude that the theoretical prediction put forward in Eq. (\ref{eq9}) is in agreement with experimental data.  A more detailed comparison with random matrix based computations over a broad range of uniform attenuation is presented in Supp. Mat. section IV \cite{SuppMat}.

We have also examined the experimentally obtained statistics of $\text{Im}[\tau_\text{W}]$.  In this case the distribution is found to be symmetric about $0$ to very good approximation. As seen in the insets of Fig. \ref{fig3} (a) and (b), we find that the mean of $\text{Im}[\tau_\text{W}]$ is consistent with theoretically predicted zero value for all levels of uniform attenuation in the graphs. 

We now turn out attention back to the power-law tails for the distributions of $\text{Re}[\tau_\text{W}]$ and $\text{Im}[\tau_\text{W}]$ presented in Figs. \ref{fig1} and \ref{fig2}.  Theory discussed in Supp. Mat. section II \cite{SuppMat} predicts that the tails of the PDFs will behave as $\mathcal{P}(\text{Re}[\tau_\text{W}])\propto 1/\text{Re}[\tau_\text{W}]^3$, on both the positive and negative sides, as long as $M\text{Re}[\tau_\text{W}]/\tau_\text{H} \gg 1/\tilde{\eta}$.  This behavior is clearly observed on the negative side of the PDF, as shown in Fig. \ref{fig1}(a).  The tail on the positive side is more complicated due to a second power-law expected in the intermediate range: $\mathcal{P}(\text{Re}[\tau_\text{W}])\propto 1/\text{Re}[\tau_\text{W}]^4$ when $1 \ll M\text{Re}[\tau_\text{W}]/\tau_\text{H} \ll 1/\tilde{\eta}$. Unfortunately we were not able to obtain such data within this range (requiring very low attenuation $\tilde{\eta}$) experimentally, but a narrow range of $\text{Re}[\tau_\text{W}]/\tau_\text{H}$ between approximately $0.3$ and $1$ in Fig. \ref{fig1}(b) shows a steeper power-law behavior, consistent with $\mathcal{P}(\text{Re}[\tau_\text{W}])\propto 1/\text{Re}[\tau_\text{W}]^4$, giving way to a more shallow slope at larger values of $\text{Re}[\tau_\text{W}]/\tau_\text{H}$, consistent with the theory.  As seen in Fig. \ref{fig2}, the distribution of the imaginary part of the time delay has a wide range with a power law  $\mathcal{P}(|\text{Im}[\tau_\text{W}]|)\propto 1/|\text{Im}[\tau_\text{W}]|^3$, consistent with our theoretical prediction.

\emph{Discussion.}  We demonstrated that the complex Wigner time delay is an experimentally accessible object sensitive to the statistics of $S$-matrix poles in the complex energy/frequency plane. In particular, we revealed theoretically and confirmed experimentally that the mean of the fluctuating $\text{Re}[\tau_\text{W}]$ is directly related to the (integrated) mean density of the imaginary parts (widths) of the resonance poles in the complex energy plane.   In addition to the experimental results discussed above, we have also employed Random Matrix Theory, as well as associated numerical simulations, for studying the distribution of the complex Wigner time delay. These simulations (Supp. Mat. section IV \cite{SuppMat}) assume a constant modal density (appropriate for a graph), but otherwise adopt pure RMT statistical properties for the system.  Through these simulations we can explore much smaller, and much larger, values of uniform attenuation than can be achieved in the experiment.  These simulations show agreement with all major predictions of the RMT-based theory, including the existence of an intermediate power-law on the positive side of the $\mathcal{P}(\text{Re}[\tau_\text{W}])$ distribution for low-loss systems.  More details and discussion about the evolution of the PDFs of complex Wigner time delay with uniform attenuation, and non-ideal coupling, can be found in Supp. Mat. section IV \cite{SuppMat}.

\emph{Conclusions.}  
We have experimentally verified the theoretical prediction that the mean value of the $\text{Re}[\tau_\text{W}]$ distribution function for a system with uniform absorption strength $\eta$ counts the fraction of scattering matrix poles with imaginary parts exceeding $\eta$.  The prediction was tested with an ensemble of microwave graphs with either one or two scattering channels, and showing broken time-reversal invariance and variable uniform attenuation.  The tails of the distributions of both real and imaginary time delay are found to agree with theory.

We acknowledge Jen-Hao Yeh for early experimental work on complex time delay statistics. This work was supported by AFOSR COE Grant No. FA9550-15-1-0171 and ONR Grant No. N000141912481. Y.V.F. acknowledges a financial support from EPSRC Grant EP/V002473/1.

\bibliography{WTD.bib}

\end{document}


\title{SUPPLEMENTARY MATERIAL for \\
Statistics of Complex Wigner Time Delays as a counter of S-matrix poles: Theory and Experiment}

\author{Lei Chen}
 \affiliation{Quantum Materials Center, Department of Physics, University of Maryland, College Park, MD 20742, USA}
 \affiliation{Department of Electrical and Computer Engineering, University of Maryland, College Park, MD 20742, USA}

\author{Steven M. Anlage}
 \affiliation{Quantum Materials Center, Department of Physics, University of Maryland, College Park, MD 20742, USA}
 \affiliation{Department of Electrical and Computer Engineering, University of Maryland, College Park, MD 20742, USA}

\author{Yan V. Fyodorov}
 \affiliation{Department of Mathematics, King’s College London, London WC26 2LS, United Kingdom}
 \affiliation{L. D. Landau Institute for Theoretical Physics, Semenova 1a, 142432 Chernogolovka, Russia}
 
\date{\today}

\maketitle

\renewcommand{\thefigure}{S\arabic{figure}}
\renewcommand{\theequation}{S\arabic{equation}}

Here we provide the reader with some additional details for the calculations described in the text of the Letter.  Section I offers a proof of Eq. (9) in the main text.  Section II discusses the tails of the distribution functions of the complex Wigner time delay.  Section III discusses the convention that we employ for the evolution of the phase of the $S$-matrix with frequency.  In Section IV we discuss the use of random matrix computations to examine the distribution functions of the complex Wigner time delay as a function of uniform attenuation.  Section V has a discussion of how the loss parameter of the graph is determined in the experiment, and how to estimate the error bars in Fig. 3 of the main text.  Section VI shows quantitative time-reversal invariance breaking effects produced by the circulator in the microwave graph system.

\section{Counting Resonance Widths via complex Wigner time delays}

Denote by $H$ the $N \times N$ Hamiltonian of the closed system, by $W$ the $N \times M$ matrix of coupling elements between the $N$ modes of $H$ and the $M$ scattering channels. The total $S$ matrix has the form:
\begin{align}
    \label{eqS1}
    \mathcal{S}(E)=1_M - 2\pi i W^{\dag} \frac{1}{E-H+i\Gamma_W} W \text{ where } \Gamma_W = \pi WW^{\dag}
\end{align}
Note that the $S$-matrix poles $\mathcal{E}_n = E_n -i\Gamma_n$ (with $\Gamma_n>0$) are eigenvalues of $H-i\Gamma_W$.

In the presence of uniform absorption with strength $\eta$, the $S$ matrix is evaluated at complex energy $S(E + i\eta) := S_{\eta}(E)$. The determinant of $S_{\eta}(E)$ is then:
\begin{align}
    \label{eqS2}
    \det S_{\eta}(E) &\coloneqq \det S(E+i\eta) \\
    \label{eqS3}
    &= \frac{\det [E-H+i(\eta  - \Gamma_{W})]}
    {\det [E-H+i(\eta  + \Gamma_{W})]} \\
    \label{eqS4}
    &= \prod_{n=1}^{N} \frac{E+i\eta - \mathcal{E}_n^*}{E+i\eta - \mathcal{E}_n},
\end{align}

Extending the definition of the Wigner time delay to uniformly absorbing systems as 
\begin{align}
    \label{eqS5}
    \tau_\text{W}(E;\eta) \coloneqq \frac{-i}{M} \frac{\partial}{\partial E} \log \det S_{\eta}(E)
\end{align}

we now have a complex quantity
\begin{align}
    \label{eqS6}
    \tau_\text{W}(E;\eta) = -\frac{i}{M} \sum_{n=1}^{N} \left( \frac{1}{E+i\eta-E_n-i\Gamma_n} - \frac{1}{E+i\eta-E_n+i\Gamma_n} \right)
\end{align}

whose real and imaginary part is given by:
\begin{align}
    \label{eqS7}
    \text{Re}\ \tau_\text{W}(E;\eta) &= \frac{1}{M} \sum_{n=1}^{N} \left[ \frac{\Gamma_n + \eta}{(E-E_n)^2 + (\Gamma_n + \eta)^2} - \frac{\eta -\Gamma_n }{(E-E_n)^2 + (\Gamma_n - \eta)^2} \right], \\
    \label{eqS8}
    \text{Im}\ \tau_\text{W}(E;\eta) &= -\frac{1}{M} \sum_{n=1}^{N} \left[ \frac{4\eta \Gamma_n(E - E_n)}{[(E-E_n)^2 + (\Gamma_n - \eta)^2] [(E-E_n)^2 + (\Gamma_n + \eta)^2]}  \right]
\end{align}

When the $S$-matrix is unitary, i.e. $\eta = 0$, the time delay is purely real and reduces to conventional Wigner time delay:
\begin{align}
    \label{eqS9}
    \tau_\text{W}(E;0) = \frac{1}{M} \sum_{n=1}^{N} \frac{2\Gamma_n}{(E-E_n)^2 + \Gamma_n^2} \coloneqq \tau_\text{W}(E)
\end{align}

All the equations above are valid for arbitrary $\eta$. There are two characteristic energy scales in the system for energies around a value $E$. First is the \textit{microscopic} one, the mean spacing between $E_n$ in the `closed' counterpart of our scattering system $\Delta = 1/(N\nu(E))$ where $\nu(E) = \frac{1}{N} \langle \sum_{n=1}^{N} \delta(E-E_n) \rangle$ is the mean density of resonance positions (in the case of Random Matrix Theory (RMT) the latter is the Wigner semicircle $\nu(E) = \frac{1}{2\pi} \sqrt{4-E^2}$). A second scale $J$ is \textit{macroscopic} and reflects a characteristic scale on which the mean density substantially changes (in RMT it is simply the width of the semicircle, $J \sim 1$). We will also introduce a useful notion of \textit{mesoscopic} energy intervals $I_E$ defined by $E_L < E < E_R$. Those are intervals with the length $|I| \coloneqq |E_R-E_L|$ satisfying $\Delta \ll |I| \ll J$. In other words, they contain a lot of resonances inside, but the density of those resonances along the real axis can be assumed to be constant. Correspondingly, we will introduce the notion of the \textit{mesoscopic energy average}, defined for any energy-dependent function $f(E)$ as
\begin{align}
    \label{eqS10}
    \langle f(E) \rangle_E = \frac{1}{|I|} \int_{E_L}^{E_R} f(E) \,dE
\end{align}

We will be interested in situations when both the typical resonance widths $\Gamma_n$ and the absorption parameter $\eta$ are of the order of the microscopic scale $\Delta$ (which does not necessarily mean that the resonances are isolated: some $\Gamma_n$ can be several times larger than $\Delta$, but they are considered to be always smaller than any \textit{mesoscopic} scale). The above situation is always typical as long as the number of open channels $M$ is of the order of unity ($M = 1$ and $M = 2$ for example). In such a situation no more than $M$ (out of $N$) resonances can violate the above condition.

Our main statement is the following: under the above assumptions the mesoscopic energy average of $\text{Re}[\tau_\text{W}(E;\eta)]$ is given by
\begin{align}
    \label{eqS11}
    \langle \text{Re}[\tau_\text{W}(E;\eta)] \rangle_E = \frac{2\pi}{M\Delta} \times \text{Prob} (\text{resonance\ widths} > \eta)
\end{align}
where we defined
\begin{align*}
    \text{Prob} (\text{resonance\ widths} > \eta) \coloneqq \frac{\#[\Gamma_n>\eta\ \text{such\ that}\ E_n\ \text{is\ inside}\ I_E]}{\text{total\ \#\ resonances\ inside}\ I_E}
\end{align*}

To verify the above statement we consider the integral:
\begin{align}
    \label{eqS12}
    \int_{E_L}^{E_R} \frac{\delta_n}{(E-E_n)^2 + \delta_n^2} \,dE = \text{sign} (\delta_n) \int_{(E_L-E_n)/|\delta_n|}^{(E_R-E_n)/|\delta_n|} \frac{dx}{x^2+1} \\
    = \text{sign} (\delta_n) \left\{ \arctan(\frac{E_R-E_n}{|\delta_n|}) - \arctan(\frac{E_L-E_n}{|\delta_n|}) \right\} \nonumber
\end{align}

We need to apply it to the right-hand side of Eq. (\ref{eqS7}) where $\delta_n = \eta \pm \Gamma_n$. We see that for the overwhelming majority of the summation index $n = 1, 2, \dots, N$ there simultaneously holds two strong inequalities
\begin{align*}
    \frac{|E_R - E_n|}{|\delta_n|} \gg 1 \quad \text{and} \quad 
    \frac{|E_L - E_n|}{|\delta_n|} \gg 1.
\end{align*}
Indeed, those inequalities can be violated only in the vicinity of the ends of the mesosocopic interval, i.e. when $|E_R,L - E_n| \sim \Delta$. The number of such terms is clearly of the order $\Delta/|I|$ which is a small parameter in the mesoscopic case. Neglecting those cases, we always can consider the arguments of $\arctan$ to be large in absolute value, hence to use $\arctan(a) \approx \frac{\pi}{2} \text{sign}(a) - \frac{1}{a} + \dots$. The contribution of subleading terms can be estimated separately (and indeed shown to be small, this time as $\Delta/J$), and the leading terms give:
\begin{align}
    \label{eqS13}
    \langle \text{Re}[\tau_\text{W}(E;\eta)] \rangle_E \approx \frac{\pi/2}{M|I|} \sum_{n=1}^{N} \left\{\left[ \text{sign} \left(\frac{E_R-E_n}{\eta+\Gamma_n} \right) - \text{sign} \left(\frac{E_L-E_n}{\eta+\Gamma_n} \right) \right] \right. \nonumber\\
    \left.- \left[ \text{sign} \left(\frac{E_R-E_n}{\eta-\Gamma_n} \right) - \text{sign} \left(\frac{E_L-E_n}{\eta-\Gamma_n} \right) \right] \right\} 
\end{align}
It is now evident that if $E_n$ is outside of the mesoscopic interval (that is $E_n < E_L < E_R$ or $E_n > E_R > E_L$) the corresponding terms in the sum (\ref{eqS13}) vanish, whereas inside the interval (for $E_L < E_n < E_R$) remembering $\eta + \Gamma_n > 0$ we see the corresponding terms in the summand are equal to $2(1 - \text{sign}(\eta -\Gamma_n)) = 4\theta(\Gamma_n - \eta)$ where we introduced the step function $\theta(x) = 1$ for $x > 0$ and $\theta(x) = 0$ otherwise.
\begin{align}
    \label{eqS14}
    \langle \text{Re}[\tau_\text{W}(E;\eta)] \rangle_E \approx \frac{2\pi}{M|I|} \sum_{n=1}^{N} \theta(\Gamma_n - \eta)
\end{align}
Finally, remembering that under our assumptions $\#(E_n \in I) \approx |I|/\Delta$ we arrive at the statement Eq. (9) in the main text.

\textbf{Remarks:} The mesoscopic energy average is defined in a given system and does not involve any ensemble average. Actually, we separately proved that if one employs the RMT ensemble average (which we denote with the bar below) instead of the \textit{mesoscopic energy average} the relation Eq. (9) holds even if we use $\tau_\text{W}(E;\eta)$ rather than $\text{Re}[\tau_\text{W}(E;\eta)]$, namely:
\begin{align}
    \label{eqS15}
    \overline{\tau_\text{W}(E;\eta)} = \frac{2\pi}{M\Delta} \int_{\tilde{\eta}}^{\infty} \rho_{\beta}^{(M)}(y) \,dy
\end{align}
where $\tilde{\eta} = 2\pi\eta/\Delta$ and $\rho_{\beta}^{(M)}(y)$ is the probability density of scaled resonance widths $y_n=2\pi|\Gamma_n|/\Delta$. We see that is exactly equivalent to mesoscopic energy averaging. This means that the mesoscopic average of $\text{Im}[\tau_\text{W}(E;\eta)]$ should be parametrically smaller than for $\text{Re}[\tau_\text{W}(E;\eta)]$, and tend to zero when the length of the mesoscopic interval formally tends to infinity. 

Thus, one can compare the result to known RMT expressions. In particular, for $\beta = 2$ and general two-port system one has \cite{Fyod96,Fyodorov97,FyodKhor99}:
\begin{align}
    \label{eqS16}
    \rho_{\beta=2}^{(M=2)}(y) = \frac{e^{-yg_1} - e^{-yg_2}}{g_1 - g_2} \left( g_1g_2\phi(y) - (g_1 + g_2)\frac{d\phi}{dy} + \frac{d^2\phi}{dy^2} \right)
\end{align}
where we denoted $\phi(y) = \frac{\sinh{y}}{y}$ and introduced coupling constants $g_1 \geqslant 1,\ g_2 \geqslant 1$ are determined from the mean (ensemble-averaged) scattering matrix which is in that model diagonal $\overline{S_{ab}} = \delta_{ab} \overline{S_{aa}}$. Namely:
\begin{align}
    \label{eqS17}
    |\overline{S_{ab}}|^2 = \frac{g_a-1}{g_a+1}
\end{align}
Closed channel $a$ corresponds to $g_a \to \infty$, perfect coupling to $g_a = 1$. If two channels are equivalent: $g_1 = g_2 = g$ we have a more compact formula:
\begin{align}
    \label{eqS18}
    \rho_{\beta=2}^{(M=2)}(y) = y \frac{d^2}{dy^2} \left( e^{-yg}\phi(y) \right)
\end{align}

Similar, but more complicated (still explicit, but in terms of 3-fold integrals) expressions are available for $\beta = 1$, see \cite{Somm99}.
For a single-channel GOE system a much simpler explicit formula for the resonance density has been recently derived \cite{FyoOsm21}, with only one-fold integrals involved. 

\section{Statistical Distribution of complex Wigner time delays: Tails}

Using the standard resonance representation for the unitary time delay (\ref{eqS9}) one can describes mechanisms \cite{Fyodorov97} responsible for the formation of various regimes in the far tail of the probability density for normalized Wigner time delays $t_w=\frac{\Delta}{2\pi}\tau_\text{W}$. Here we provide a similar consideration for the normalized real part: $\tilde{t}_w=M\frac{\Delta}{2\pi}\text{Re}[\tau_\text{W}]$ in the presence of a uniform absorption $\eta>0$. Inspection of the representation Eq. (\ref{eqS7}) makes it clear that anomalously high values of the time delays happen when (\textbf{i})) the observation energy value $E$ is anomalously close to $E_n$ and simultaneously (\textbf{ii})) the resonance widths $\Gamma_n$ comes anomalously close to the absorption value $\eta$, that is $\Gamma_n-\eta \ll \eta$. In such an event the second term in Eq. (\ref{eqS7}) is dominant, and therefore a faithful model for the tail formation can be obtained by considering the following approximation: 
\begin{align}
    \label{eqS19}
    \tilde{t}_w \approx \frac{\Delta}{2\pi} \frac{\Gamma_n-\eta}{(E-E_n)^2 + (\Gamma_n-\eta)^2} \equiv \frac{y-\tilde{\eta}}{x^2+(y-\tilde{\eta})^2}
\end{align}
where the scaled resonance widths $y=\frac{2\pi}{\Delta}\Gamma_n$ is distributed with the probability density $\rho_\beta^{(M)}(y)$ and the variable $x=\frac{2\pi}{\Delta}(E-E_n)$ can be considered for our purposes as uniformly distributed in the interval $[-a,a]$ where $a$ is any constant of the order of unity. We will take $a=1$ for simplicity. Using the symmetry $x \to -x$ and introducing $w=x^2$ one can write the probability density $\mathcal{P}(\tilde{t}_w)$ in this approximation as 
\begin{align}
    \label{eqS20}
    \mathcal{P}(\tilde{t}_w) = \int_0^{\infty} \rho_{\beta}^{(M)}(y) \,dy \int_0^1 \delta \left(\tilde{t}_w-\frac{y-\tilde{\eta}}{w+(y-\tilde{\eta})^2} \right) \,\frac{dw}{\sqrt{w}}
\end{align}
Solving the $\delta$-constraint we find that $w=(y-\tilde{\eta})\left(\frac{1}{\tilde{t}_w} - (y-\tilde{\eta}) \right)$. Due to the constraint $w > 0$ we see that this implies that the integral over $x$ is nonzero only for $y$ in the range $\tilde{\eta} < y < \tilde{\eta}+\frac{1}{\tilde{t}_w}$ for the right tail values $\tilde{t}_w > 0$, whereas for the left tail $\tilde{t}_w < -\tilde{\eta}^{-1}$ we have $\tilde{\eta}+\frac{1}{\tilde{t}_w} < y < \tilde{\eta}$. On the other hand it is easy to see that the upper limit constraint $w < 1$ is immaterial if we are interested in the tail $\tilde{t}_w \gg 1$, and can be replaced with $w < \infty$. Performing the integration over $w$ gives
\begin{align}
    \label{eqS21}
    \mathcal{P}(\tilde{t}_w) = \frac{1}{\tilde{t}_w^2} \int_{\tilde{\eta}}^{\tilde{\eta}+\frac{1}{\tilde{t}_w}} \rho_{\beta}^{(M)}(y) \frac{y-\tilde{\eta}}{\sqrt{(y-\tilde{\eta})(\frac{1}{\tilde{t}_w}-(y-\tilde{\eta}))}} \,dy
\end{align}
and introducing $v=(y-\tilde{\eta})\tilde{t}_w$ we finally get the right tail
\begin{align}
    \label{eqS22}
    \equiv \frac{1}{\tilde{t}_w^3} \int_0^1 \rho_{\beta}^{(M)}\left(\frac{v}{\tilde{t}_w}+\tilde{\eta}\right) \sqrt{\frac{v}{1-v}} \,dv
\end{align}
We see that the following two situations are possible. First (using $\int_0^1 \sqrt{\frac{v}{1-v}} \,dv = \frac{\pi}{2}$) we see that for any $\tilde{\eta} > 0$ the most distant right tail has a universal exponent (for any $\beta$) given by
\begin{align}
    \label{eqS23}
    \mathcal{P}(\tilde{t}_w) \approx \frac{\pi}{2} \frac{\rho_{\beta}^{(M)}(\tilde{\eta})}{\tilde{t}_w^3}, \quad \tilde{t}_w \gg \frac{1}{\tilde{\eta}}
\end{align}
However, if absorption is small: $\tilde{\eta} \ll 1$ then there exists another tail regime: $1 \ll \tilde{t}_w \ll \frac{1}{\tilde{\eta}}$ where
\begin{align}
    \label{eqS24}
    \mathcal{P}(\tilde{t}_w) \approx \frac{1}{\tilde{t}_w^3} \int_0^1 \rho_{\beta}^{(M)}\left(\frac{v}{\tilde{t}_w} \right) \sqrt{\frac{v}{1-v}} \,dv,
\end{align}
and finally using that for small argument $\rho_{\beta}^{(M)}(y \ll 1) \sim \text{const}\ y^{\frac{M\beta}{2}-1}$ we arrive at the intermediate tail:
\begin{align}
    \label{eqS25}
    \mathcal{P}(\tilde{t}_w) \approx \text{const}\ \tilde{t}_w^{\,-\frac{M\beta}{2}-2}, \quad 1 \ll \tilde{t}_w \ll \frac{1}{\tilde{\eta}}
\end{align}
In fact this tail is exactly the same as that derived in \cite{Fyodorov97,Fyodorov97a} for $\tilde{\eta} = 0$.  Note that for the $M=2$ port, $\beta = 2$ data shown in Fig. 1(b) of the main text, the power-law of the intermediate tail is expected to be $\mathcal{P}(\tilde{t}_w) \propto \tilde{t}_w^{-4}$.

Finally, for negative time delay it is easy to show that the far tail for $\tilde{t}_w < -\tilde{\eta}^{-1}$ is given by the same result (\ref{eqS23}), with $\tilde{t}_w \to |\tilde{t}_w|$, and this is the only asymptotic regime on the left ($\tilde{t}_w <0$).

Now we study the far tails of the $J_w=-M\text{Im}[\tau_\text{W}]/\tau_\text{H}$ which in the same approximation can be extracted from (\ref{eqS8}) as 
\begin{align}
    \label{eqJA}
    J_w \approx  \frac{4\tilde{\eta}yx}{[x^2+(y-\tilde{\eta})^2][x^2+4\tilde{\eta}^2]}\approx \frac{yx}{\tilde{\eta}[x^2+(y-\tilde{\eta})^2]}
\end{align}
where we used that the far tail values $|J_w|\gg 1/\tilde{\eta}$ come when $x\ll \tilde{\eta}$. Hence we also can safely consider $-\infty<x<\infty$ and
write 
the probability density $\mathcal{P}(\tilde{t}_w)$ in this approximation as 
\begin{align}
    \label{eqJB}
    \mathcal{P}\left(|J_w|\gg \tilde{\eta}^{-1}\right) = \int_0^{\infty} \rho_{\beta}^{(M)}(y) \,dy \int_{-\infty}^{\infty} \delta \left(J_w-\frac{1}{\tilde{\eta}}\frac{yx}{x^2+(y-\tilde{\eta})^2} \right) \,dx
\end{align}
Note that such a density is symmetric: $\mathcal{P}\left(J_w\right)= \mathcal{P}\left(-J_w\right)$, so we consider $J_w>0$. Solving the delta-functional constraint for $x$, we find two values of $x$ contributing: 
\begin{align}
    \label{eqJC}
    x_{1,2}=\frac{1}{2}\left(\frac{y}{J_w\tilde{\eta}}\mp \sqrt{\left(4-\frac{1}
    {J_w^2\tilde{\eta}^2}\right)(y-y_{+})(y_{-}-y)}\right)
\end{align}
as long as $y_{+}<y<y_{-}$ where we defined 
\begin{align}
    \label{eqJC1}
    y_{\pm}=\frac{\tilde{\eta}}{1\pm \frac{1}{2J_w\tilde{\eta}}}  
\end{align}
This gives 
\begin{align}
    \label{eqJD}
    \mathcal{P}\left(|J_w|\gg \tilde{\eta}^{-1}\right) = \frac{1}{2}\int_{y_{+}}^{y_{-}} \rho_{\beta}^{(M)}(y)   \left(\frac{1}{|\phi'(x_1)|}+\frac{1}{|\phi'(x_2)|} \right) \,dy, \quad \phi(x):=\frac{1}{\tilde{\eta}}\frac{yx}{x^2+(y-\tilde{\eta})^2}
\end{align}
Note that for $J_w\tilde{\eta}\gg 1$ the width of the integration domain over $y$ is much smaller than the typical values $y\sim \tilde{\eta}$ as $y_{-}-y_{+}\approx\frac{1}{J_w}\ll \tilde{\eta}$. Using this and exploiting the relation $J=\phi(x_{1,2})$ we can approximate 
\[
\frac{1}{|\phi'(x_{1,2})|}\approx \frac{1}{J_w^2}\frac{x_{1,2}^2}{\left|(y-\tilde{\eta})^2-x_{1,2}^2\right|}
\]
and in this way arrive to:
\begin{align}
    \label{eqJE}
    \mathcal{P}\left(J_w\gg \tilde{\eta}^{-1}\right) \approx \frac{\rho_{\beta}^{(M)}(\tilde{\eta})}{2J_w^2}
    \left(I_1+I_2\right), \quad I_{1,2}=\int_{y_{+}}^{y_{-}} \frac{x_{1,2}^2}{\left|(y-\tilde{\eta})^2-x_{1,2}^2\right|} \,dy \quad 
\end{align}
where $x_{1,2}\approx \frac{y}{2J_w\tilde{\eta}}\pm \sqrt{(y-y_{+})(y_{-}-y)}$. Evaluation of the two integrals goes in a similar way, so we consider only 
\[
I_{1}=\int_{y_{+}}^{y_{-}} \frac{\left(\frac{y}{2J_w\tilde{\eta}}+\sqrt{(y-y_{+})(y_{-}-y)}\right)^2}{\left|\left(y-\tilde{\eta}-\frac{y}{2J_w\tilde{\eta}}-\sqrt{(y-y_{+})(y_{-}-y)}\right)\left(y-\tilde{\eta}+\frac{y}{2J_w\tilde{\eta}}+\sqrt{(y-y_{+})(y_{-}-y)}\right)\right|} \,dy
\]
We first change variables as $y=y_{+}+(y_{-}-y_{+})t, \, 0<t<1$
and use that for $J_w\tilde{\eta}\gg 1$ we can write
\[
\frac{y_{+}}{2J_w\tilde{\eta}}\approx \frac{1}{2J_w}, \quad y_{+}-\tilde{\eta}-\frac{y}{2J_w\tilde{\eta}}\approx 0, \quad y_{+}-\tilde{\eta}+\frac{y}{2J_w\tilde{\eta}}\approx \frac{1}{J_w},  \quad  y_{-}-y_{+}\approx\frac{1}{J_w}
\]
Applying the above systematically and keeping only the leading order one finds after further algebraic manipulations that
\[
I_{1}\approx \frac{1}{J_w} \int_{0}^{1} \frac{\left(\frac{1}{2}+\sqrt{t(1-t)}\right)^2}{\sqrt{t(1-t}
\left(\sqrt{t}+\sqrt{1-t}\right)^2} \,dt
\]
The integral is well-defined and convergent and yields some positive constant whose value is however immaterial for us (in fact, substituting $t=\sin^2{\alpha}, \, \alpha\in(0,\pi/2)$ brings it to a nice form). We therefore conclude that asymptotiucally both $I_1$ and $I_2$ are proportional to the factor $J_w^{-1}$ which finally implies the tail formula:
\begin{align}
    \label{eqJE1}
    \mathcal{P}\left(|J_w|\gg \tilde{\eta}^{-1}\right) \approx \mbox{const}\times \frac{\rho_{\beta}^{(M)}(\tilde{\eta})}{2J_w^3}
\end{align}

\section{Sign Convention for the Phase Evolution of the $S$-Matrix Elements}
It should be noted that there are two widely-used conventions for the evolution of the phase of the complex $S$-matrix elements with increasing frequency. Microwave network analyzers utilize a convention in which the phase of the scattering matrix elements \emph{decreases} with increasing frequency. Here we adopt the convention used in the theoretical literature that the phase of $S$-matrix elements \emph{increases} with increasing frequency.

\section{Random Matrix Theory Simulation and Time Delay Distributions}
In this section, we utilize numerical data from the Random Matrix Theory (RMT) simulation to further examine the theory presented in this paper, and provide more insights for discussion. The RMT data is generated using Random Matrix Monte Carlo simulation \cite{Hemmady2006Thesis}.

\begin{figure}[ht]
\includegraphics[width=\textwidth]{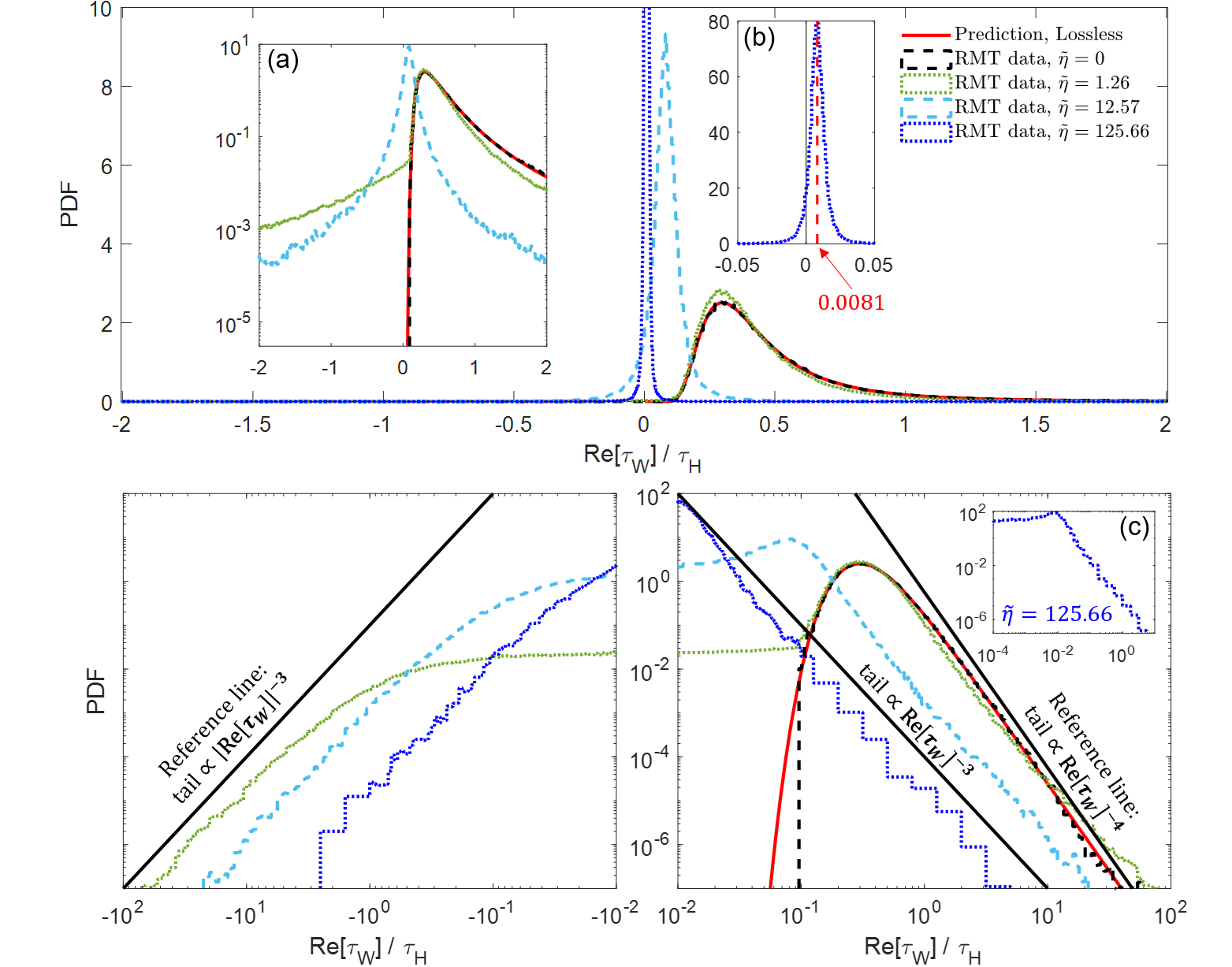}
\caption{Evolution of the PDF of simulated $\text{Re}[\tau_\text{W}]$ with increasing uniform attenuation ($\tilde{\eta}$) from an ensemble of two-port ($M=2$) GUE ($\beta=2$) RMT numerical data. The upper figure is the linear-linear plot of the distribution of $\text{Re}[\tau_\text{W}]$, while the lower one is the log-log version of the same data. Inset (a) and (b) show the zoom-in view of the PDFs for different attenuation values, and the mean value of $\text{Re}[\tau_\text{W}]$ is 0.0081 at $\tilde{\eta}=125.66$. Inset (c) shows the whole PDF of the positive $\text{Re}[\tau_\text{W}]$ in log-log scale for $\tilde{\eta}=125.66$. The reference lines are added in the log-log plot to characterize the power-law tail features of the PDFs.}
\label{figS1}
\end{figure}

\begin{figure}[ht]
\includegraphics[width=\textwidth]{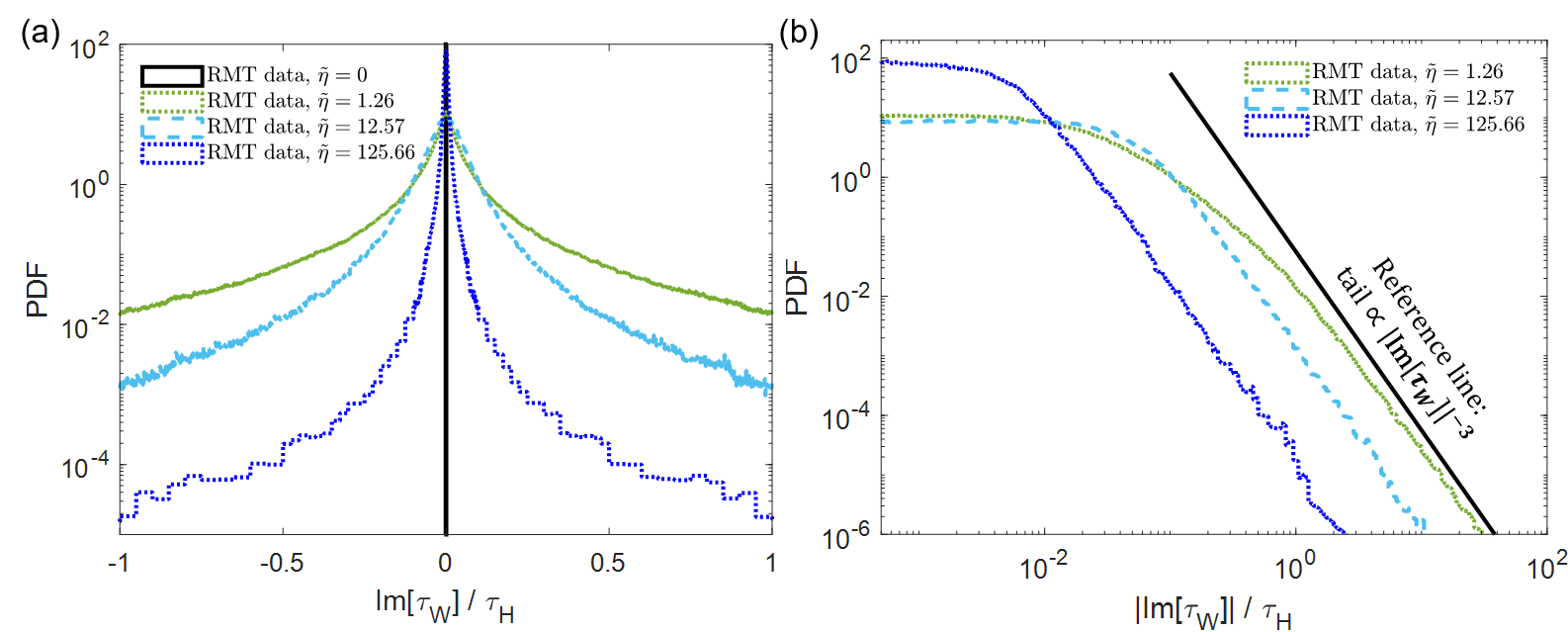}
\caption{Evolution of the PDF of simulated $\text{Im}[\tau_\text{W}]$ with increasing uniform attenuation ($\tilde{\eta}$) from an ensemble of two-port ($M=2$) GUE RMT data. (a) shows the PDFs of $\text{Im}[\tau_\text{W}]$ in a log-linear scale, while (b) shows the PDFs of $|\text{Im}[\tau_\text{W}]|$ in a log-log scale. The reference lines are added in the log-log plot to characterize the power-law tail feature of the PDFs.}
\label{figS2}
\end{figure}

Figs. \ref{figS1} and \ref{figS2} show the evolution of the PDF of simulated complex Wigner time delay $\text{Re}[\tau_\text{W}]$ and $\text{Im}[\tau_\text{W}]$ with increasing uniform attenuation ($\tilde{\eta}$) from an ensemble of GUE RMT numerical data, respectively. The upper figure in Fig. \ref{figS1} is the linear-linear plot of the PDFs, while the lower figure shows the log-log plot of the PDFs. The zoom-in view in Fig. \ref{figS1}(a) shows the detailed evolution of PDF of $\text{Re}[\tau_\text{W}]$ as the uniform attenuation increases, while Fig. \ref{figS1}(b) shows the distribution of $\text{Re}[\tau_\text{W}]$ will concentrate around its mean value (0.0081) at a large $\tilde{\eta}$ setting (strong uniform attenuation in the system). Figure \ref{figS1} shows that the peak of the PDF shifts to lower $\text{Re}[\tau_\text{W}]$ values as the uniform attenuation increases, and $\text{Re}[\tau_\text{W}]$ starts to acquire negative values -- the same behavior we have seen in the main text from the experiment. Both positive and negative sides of the PDF have a power-law tail in the log-log view of Fig. \ref{figS1}. When the uniform attenuation $\tilde{\eta}$ is zero or small, we have $\mathcal{P}(\text{Re}[\tau_\text{W}])\propto 1/\text{Re}[\tau_\text{W}]^4$ for the tail on the positive side; and as soon as the attenuation increases, the tail distribution becomes $\mathcal{P}(\text{Re}[\tau_\text{W}])\propto 1/\text{Re}[\tau_\text{W}]^3$, consistent with the theory in section II. The negative side of the PDFs always show a power-law tail of $\mathcal{P}(\text{Re}[\tau_\text{W}])\propto 1/\text{Re}[\tau_\text{W}]^3$.

\begin{figure}[!!ht]
\includegraphics[width=\textwidth]{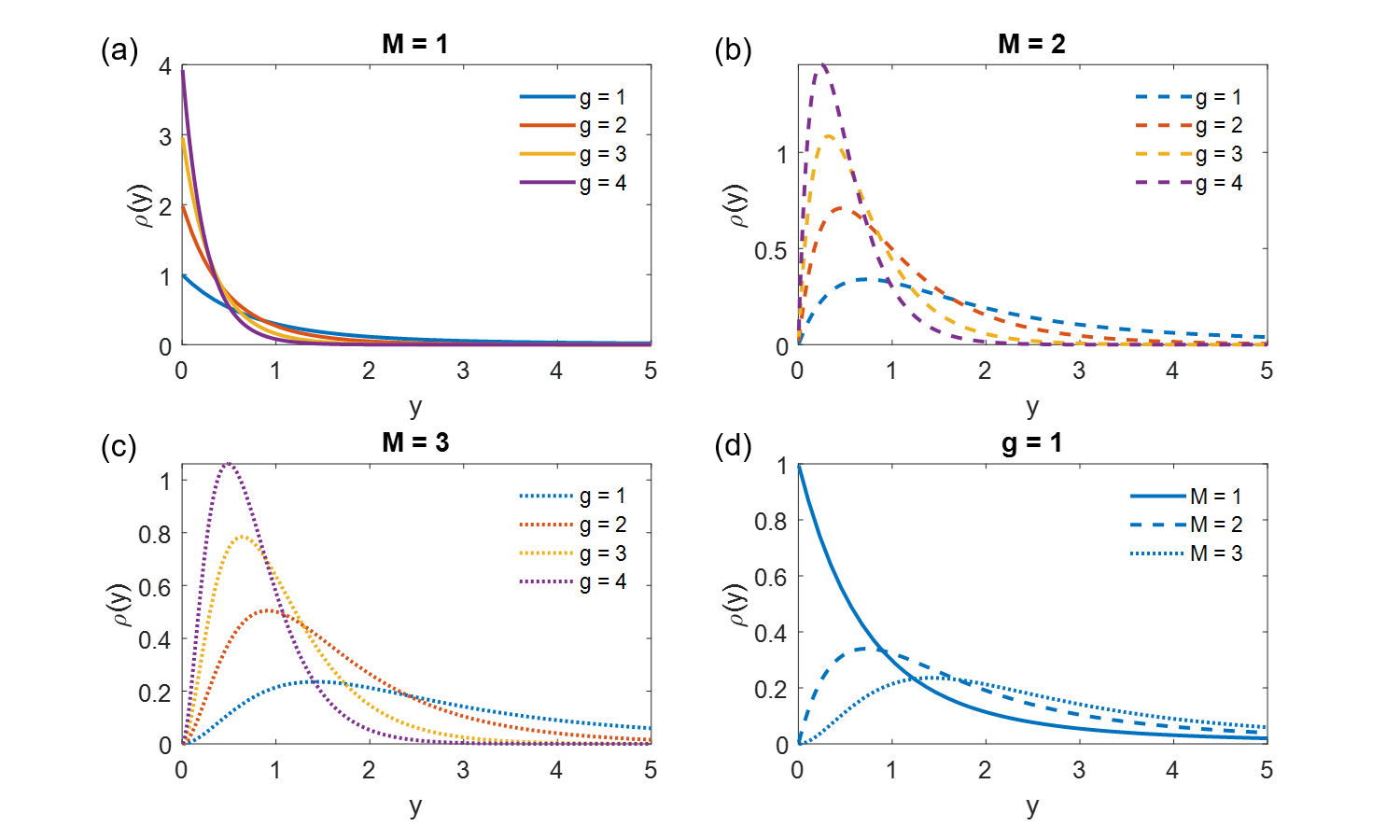}
\caption{Probability distributions $\rho(y)$ of scaled resonance width $y$ ($y=\pi\Gamma_n/\Delta$) for different numbers of scattering channels ($M$) and variable coupling strength ($g$) in the GUE lossless setting. Panels (a)--(c) show the probability distributions of the scaled resonance width with different coupling settings ($g=1,\ 2,\ 3\ \text{and}\ 4$) for $M=1,\ 2,\ \text{and}\ 3$, respectively. (d) shows the comparison between the probability distributions for different numbers of scattering channels ($M=1,\ 2,\ \text{and}\ 3$) at perfect coupling setting ($g=1$).}
\label{figS3}
\end{figure}

\begin{figure}[ht]
\includegraphics[width=0.6\textwidth]{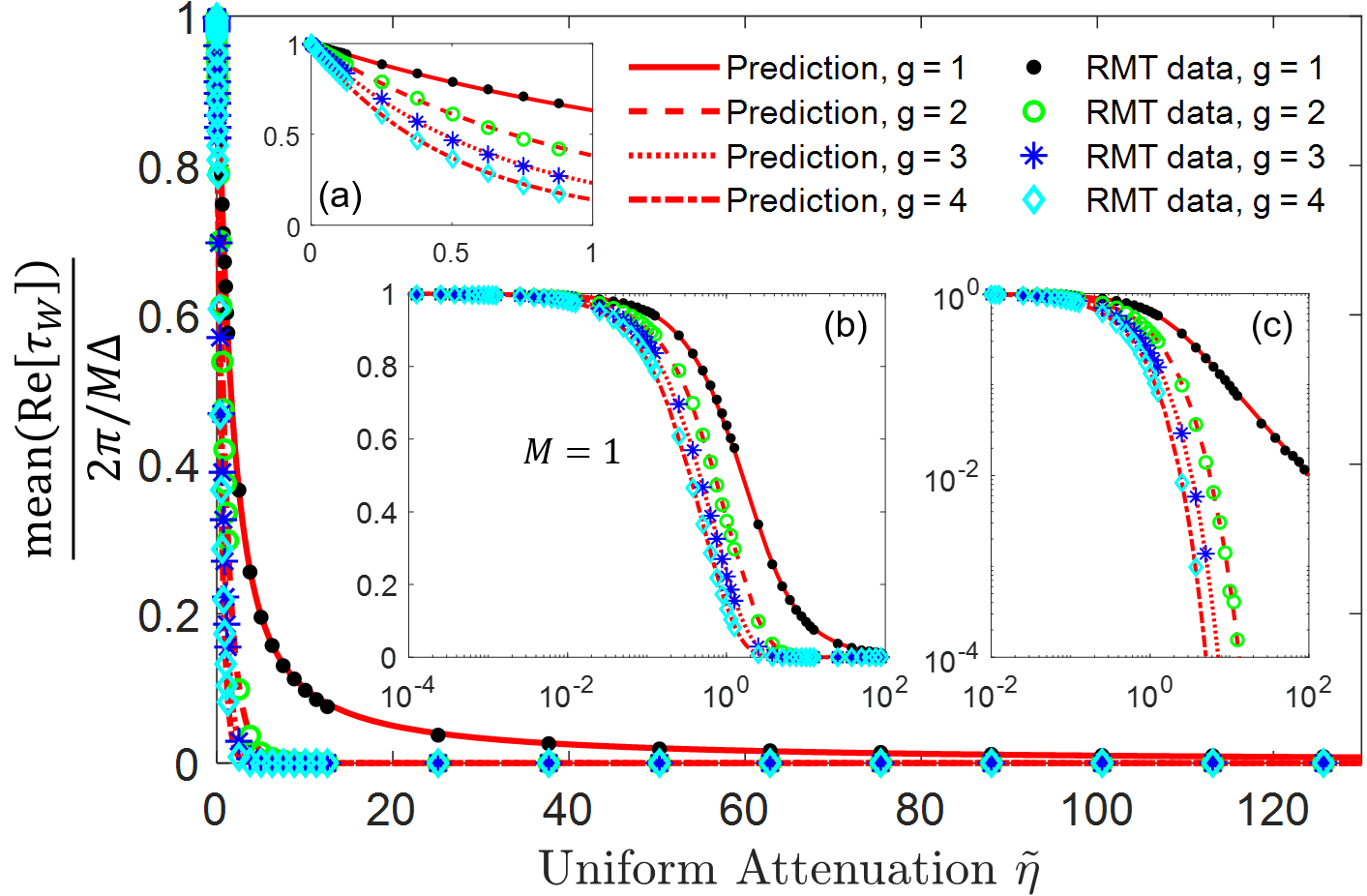}
\caption{Mean of simulated $\text{Re}[\tau_\text{W}]$ as a function of uniform attenuation $\tilde{\eta}$ with variable coupling strength ($g$) evaluated using ensembles of one-port ($M=1$) GUE RMT numerical data. The markers are RMT data, while the red lines are theoretical predictions. Inset (a) shows the zoom-in details of the plot at small attenuation values. Inset (b) and (c) are the linear-log scale and log-log scale of the plot, respectively.}
\label{figS4}
\end{figure}

\begin{figure}[ht]
\includegraphics[width=0.6\textwidth]{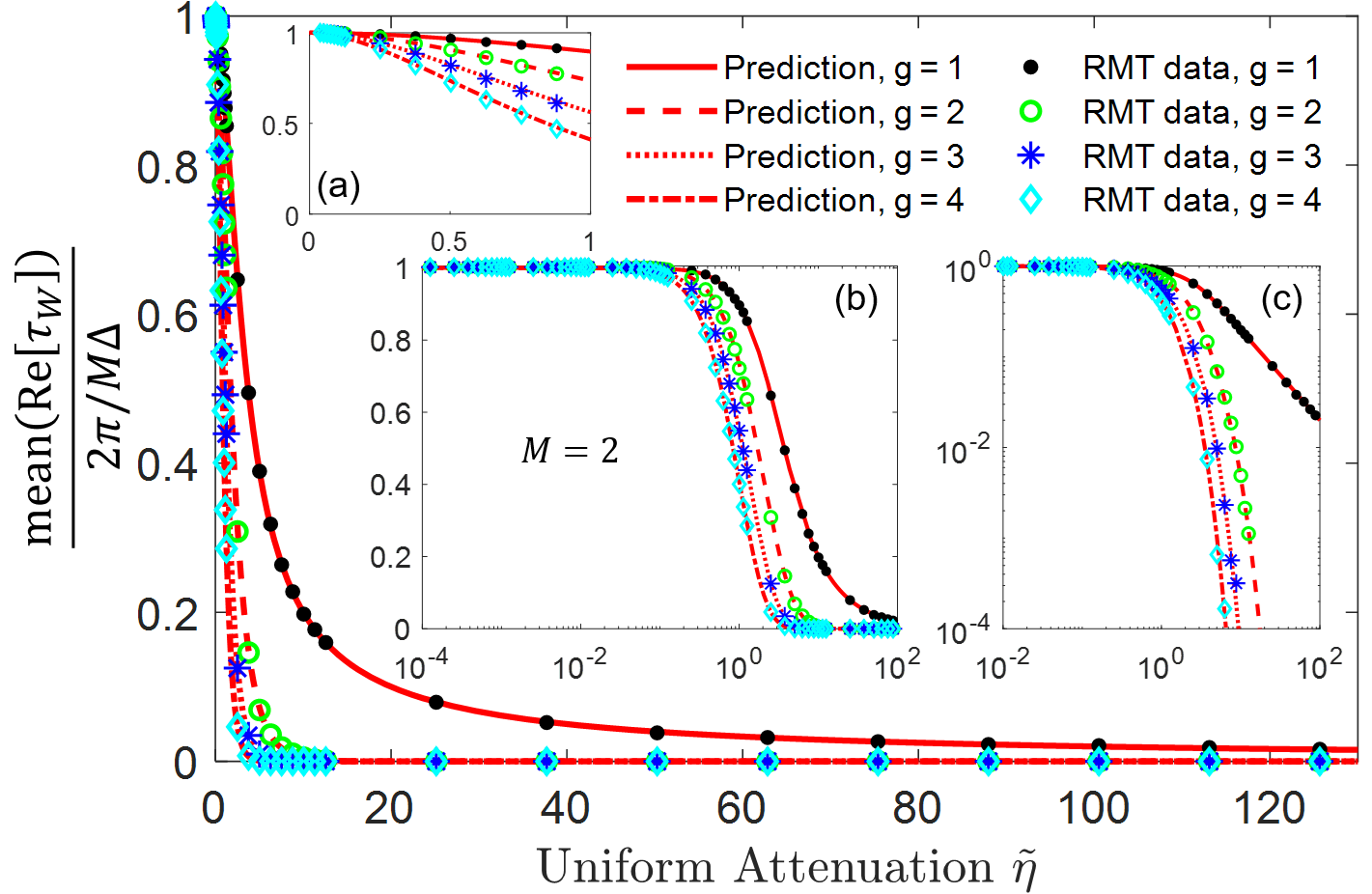}
\caption{Mean of simulated $\text{Re}[\tau_\text{W}]$ as a function of uniform attenuation $\tilde{\eta}$ with variable coupling strength ($g$) evaluated using ensembles of two-port ($M=2$) GUE RMT data. The markers are RMT data, while the red lines are theoretical predictions. Inset (a) shows the zoom-in details of the plot at small attenuation values. Inset (b) and (c) are the linear-log scale and log-log scale of the plot, respectively.}
\label{figS5}
\end{figure}

Fig. \ref{figS2}(a) shows the log-linear plot of the PDFs of $\text{Im}[\tau_\text{W}]$, while Fig. \ref{figS2}(b) shows the log-log plot of the PDFs of $|\text{Im}[\tau_\text{W}]|$ (the distributions of $\text{Im}[\tau_\text{W}]$ are symmetrical on the positive and negative sides). In Fig. \ref{figS2}(a), the PDF starts from a $\delta$-function in the lossless case, and it expands and then shrinks around the peak value (0) as $\tilde{\eta}$ increases. Fig. \ref{figS2}(b) shows the power-law tail feature of the PDF, and reference lines are added which is consistent with the theory prediction in in section II.

We also demonstrate the correctness of the theory for variable coupling settings using the RMT simulation. 
Fig. \ref{figS3} shows the probability distributions of the resonance width $\Gamma_n$ for different numbers of scattering channels ($M$) and variable coupling strength ($g$) in the GUE lossless setting, where $y=\pi\Gamma_n/\Delta$ is the scaled resonance width. Panels (a)--(c) demonstrates that the peak of the $\rho(y)$ distribution shifts to lower values as $g$ goes up, which indicates that the majority of the poles of the $S$-matrix are closer to the real axis in the lossless case when the coupling becomes weaker. Fig. \ref{figS3}(d) clearly demonstrates that the one-port ($M=1$) case is very different from the other multi-port cases.
Figs. \ref{figS4} and \ref{figS5} examine the theory further using ensembles of one-port ($M=1$) and two-port ($M=2$) GUE RMT data of variable uniform attenuation ($\tilde{\eta}$) with different coupling settings ($g$), respectively. The RMT data results are directly compared to the theory predictions calculated using the probability distribution functions shown in Fig. \ref{figS3}, and they agree quite well.



\section{Estimation of Loss parameter $\alpha$ and Error Bars}
\begin{figure}[ht]
\includegraphics[width=\textwidth]{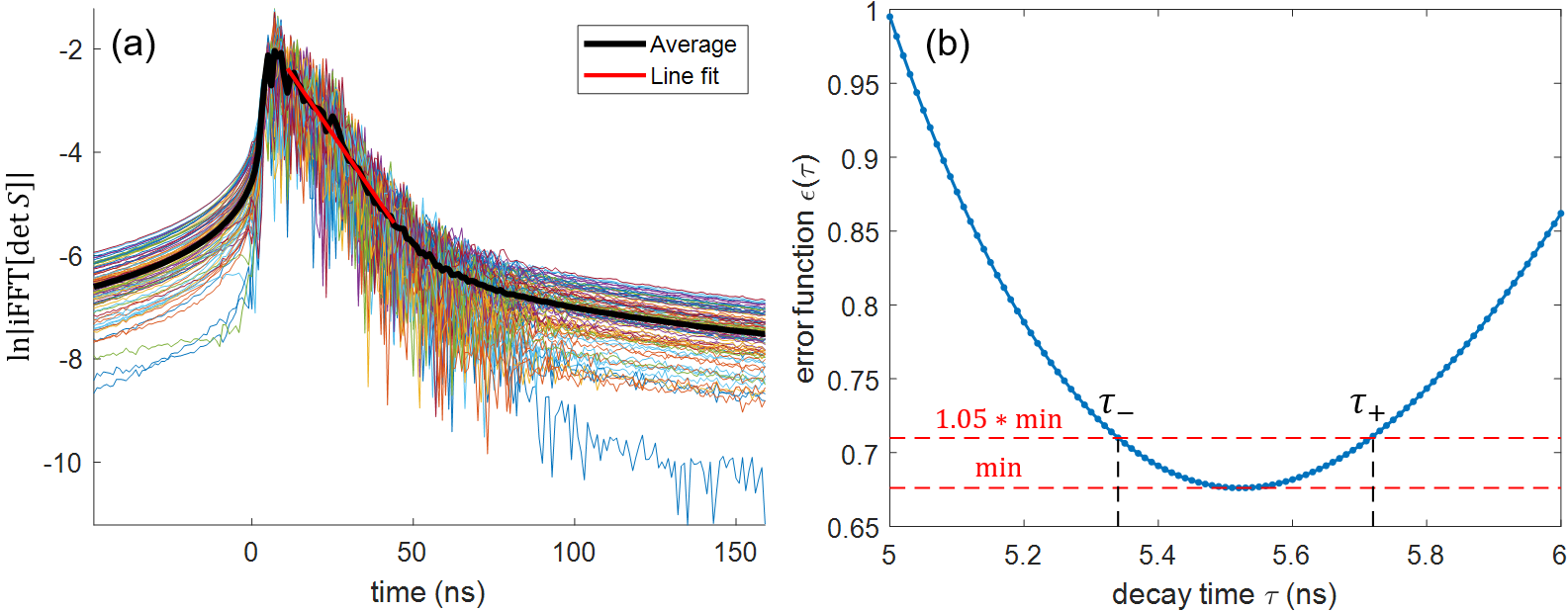}
\caption{(a) shows the fitting process of the inverse Fourier transformed $\det[S]$ data to the time domain. Multi-color lines show the data from each realization, and the black line is the average of all realizations. The red line shows the linear fit. (b) shows the error bar estimation for the decay time $\tau$. Blue dotted line shows the error function $\epsilon(\tau)$ vs the decay time $\tau$. The lower red dashed line shows the minimum level of the error function, and the upper red dashed line shows the $1.05 \times \text{minimum\ level}$. The cross points of the upper red dashed line with the blue line give the error bar $[\tau_-,\tau_+]$ for the decay time $\tau$.}
\label{figS7}
\end{figure}

In Fig. 3 of the main text, we plot the data points for the mean of the $\text{Re}[\tau_\text{W}]$ vs loss with error bars. 
The vertical error bars are determined by the statistical binning error $\sigma \sim \frac{1}{\sqrt{N_{\text{ensemble}} \times N_{\text{mode}}}}$, where $N_{\text{ensemble}}$ is the number of realizations in one ensemble, and $N_{\text{mode}}$ is the number of resonant modes in one realization, such that $N_{\text{ensemble}} \times N_{\text{mode}}$ is the total number of modes studied in one ensemble data set.
The horizontal error bar is estimated from the fitting process in calculation of the system loss parameter $\alpha$. The loss parameter $\alpha$ is defined as the ratio of the typical 3-dB bandwidth of the resonant modes to the mean mode-spacing, and it can be written as $\alpha = \frac{L_e}{2\pi c\tau}$ in the case of graph systems, where $L_e$ is the total electrical length of the graph, $c$ is the speed of light in vacuum, and $\tau$ is the energy decay time for the system. The energy decay time $\tau$ is obtained from the power decay profile (see Fig. \ref{figS7}(a)) by inverse Fourier transforming the RCM-normalized measured data for $\det[S]$ to the time domain. By fitting to the linear portion of the ensemble average power decay profile (black line), one can get the slope and the decay time $\tau$ can be computed by $\tau = -1/(2*\text{slope})$. Fig. \ref{figS7}(b) shows the estimation of error bars for the decay time $\tau$. The fitting process in Fig. \ref{figS7}(a) gives the sample dataset $(x_i,\ y_i)$, $i=1,2,...,N$ and linear function $y=kx+b$ for extracting the decay time $\tau$. Here we define an error function $\epsilon(k) = \min \left\{ \sum_{i} \left( y_i-(kx_i+b) \right)^2 \right\}$. It is easy to prove that $\epsilon(k) = \sum_i (y_i-kx_i)^2 - \frac{1}{N} \left( \sum_i (y_i-kx_i) \right)^2 $. By varying the decay time $\tau$, we can get different values of the slope $k$ and plot the error function $\epsilon(\tau)$ as a function of the decay time $\tau$ (see Fig. \ref{figS7}(b)). The minimum error function determines the best decay time $\tau$ and we use an error level of 1.05 to estimate the error bar $[\tau_-,\tau_+]$ of decay time $\tau$. The error bars of the decay time $\tau$ will then be transferred to the attenuation parameter $\tilde{\eta} = 4\pi\alpha = \frac{2L_e}{c\tau}$, and plotted as the horizontal error bars in Fig. 3 in the main text.

\section{Time-Reversal Invariance Breaking in Graphs by Microwave Circulator}

\begin{figure}[ht]
\includegraphics[width=0.6\textwidth]{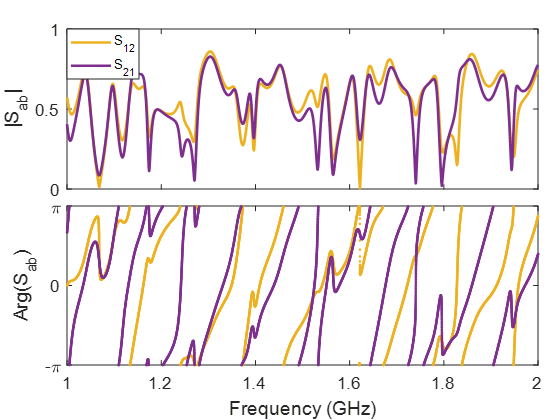}
\caption{Figure shows differences between $S_{12}$ (yellow line) and $S_{21}$ (purple line) vs frequency in a tetrahedral microwave graph containing a circulator on one internal node of the graph. In the working frequency range ($1-2$ GHz) of the microwave circulator, the two transmission parameters do not agree, neither in amplitude (upper plot) nor in phase (lower plot).}
\label{figS8}
\end{figure}

We introduce microwave circulators to the graph experiments to break the time-reversal invariance of the system \cite{awniczak2010}. From the schematic insets of Fig. 3 in the main text, we have one internal node of the graph being replaced by a microwave circulator. This non-reciprocal device brings differences to the two transmission ($S_{12} \& S_{21}$) parameters of the system, which is demonstrated in Fig. \ref{figS8}. In order to quantitatively evaluate the degree of time-reversal invariance breaking, we use the definition of time-forward and time-reversed transmission asymmetry \cite{Schafer2009Thesis} to perform the analysis: 
\begin{align}
    \label{EqAsym}
    \tilde{a} = \frac{S_{12}-S_{21}}{|S_{12}|+|S_{21}|}
\end{align}

\begin{figure}[ht]
\includegraphics[width=0.6\textwidth]{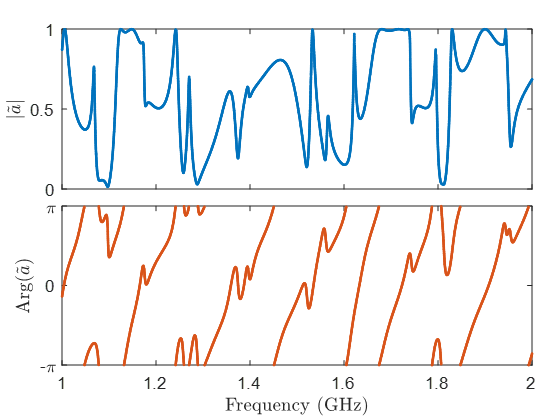}
\caption{Figure shows the time-reversal transmission asymmetry function $\tilde{a}$ vs frequency in a microwave graph with circulator ($1-2$ GHz). Upper plot shows the magnitude of $\tilde{a}$ vs frequency, and lower plot shows the phase of $\tilde{a}$ vs frequency.}
\label{figS9}
\end{figure}

This function has an absolute value from 0 (no symmetry breaking) to 1 (maximum symmetry breaking). Fig. \ref{figS9} shows an example of the asymmetry function analysis on experimental data from a realization of the tetrahedral microwave graph ($M=2$) with circulator. The asymmetry $\tilde{a}$ shows strong fluctuations as a function of frequency, but the magnitude of $\tilde{a}$ is close to 1 for many of the frequencies. The asymmetry plot in other frequency ranges shows similar behaviors. It is then well demonstrated that one circulator in such a graph setup has a satisfactory time-reversal invariance breaking effect.

\bibliography{WTD.bib}